\documentstyle[11pt,newpasp,twoside,epsf]{article}
\reversemarginpar

\begin{document}
\title{The Parkes Multibeam Pulsar Survey}
\author{F.~Camilo$^1$, A.~G.~Lyne$^1$, R.~N.~Manchester$^2$,
J.~F.~Bell$^2$, V.~M.~Kaspi$^3$, N.~D'Amico$^4$, N.~P.~F.~McKay$^1$,
F.~Crawford$^3$, I.~H.~Stairs$^1$, D.~J.~Morris$^1$,
D.~C.~Sheppard$^1$, A.~Possenti$^4$}
\affil{$^1$U. of Manchester, Jodrell Bank Observatory, Cheshire,
SK11~9DL, UK}
\affil{$^2$ATNF, CSIRO, P.O.~Box~76, Epping NSW~1710, Australia}
\affil{$^3$Center for Space Research, MIT, Cambridge, MA~02139, USA}
\affil{$^4$Osservatorio Astronomico di Bologna, 40127~Bologna, Italy}

\begin{abstract}
The Parkes multibeam pulsar survey uses a 13-element receiver operating
at a wavelength of 20\,cm to survey the inner Galactic plane with
remarkable sensitivity.  To date we have collected and analyzed data
from 45\% of the survey region ($|b| < 5^\circ$; $260^\circ < l <
50^\circ$), and have discovered 440 pulsars, in addition to
re-detecting 190 previously known ones.  Most of the newly discovered
pulsars are at great distances, as inferred from a median dispersion
measure (DM) of 400\,cm$^{-3}$\,pc.

\end{abstract}

\section{Introduction}

Pulsars are steep-spectrum radio sources, with a ``typical'' spectral
index of $-1.6$ (Lorimer et al. 1995).  For this reason, most
large-area pulsar surveys have been done at relatively low frequencies,
$\nu \sim 400$\,MHz.  However, at low frequencies and low Galactic
latitudes the contribution from synchrotron radiation (with spectral
index $\sim -2.7$) dominates the system temperature of a radio
telescope, greatly reducing the sensitivity to most pulsars.

To search for pulsars along the disk of the Galaxy, one should
therefore consider using a relatively high frequency, $\nu \sim
1400$\,MHz ($\lambda 20$\,cm).  To search for {\em distant\/} pulsars
along the Galactic plane, one is virtually compelled to use high
frequencies, because multi-path propagation of radio pulses through the
inhomogeneous interstellar medium results in broadening
(``scattering'') of intrinsically sharp pulses (see Fig.~4{\em b\/}).
This effect, greatly reducing the detectability of pulsars, varies with
frequency approximately as $\nu^{-4.4}$ (Cordes, Weisberg, \& Boriakoff
1985).  The obvious drawback of a survey at 1400\,MHz is that, in
addition to the long individual integration times required to maintain
high sensitivity due to the reduced pulsar fluxes, the number of
independent telescope pointings needed increases as $\nu^2$.

\begin{table}
\caption{Three 20\,cm pulsar surveys}
\begin{tabular}{llll}
\tableline
                               & Jodrell Bank     & Parkes      & Parkes     \\
\tableline
Latitude range, $|b|$\dotfill  & $<1^\circ$       & $<4^\circ$  & $<5^\circ$ \\
Longitude range, $l$\dotfill   & $-5^\circ$\dots$100^\circ$
                                 & $-90^\circ$\dots$20^\circ$
                                 & $-100^\circ$\dots$50^\circ$               \\
Center frequency (MHz)\dotfill & 1400             & 1520        & 1374       \\
Number of beams\dotfill        & 1                & 1           & 13         \\
Integration time (min)\dotfill & 10               & 2.5         & 35         \\
Sample interval (ms)\dotfill   & 2.0              & 1.2         & 0.25       \\
Bandwidth (MHz)\dotfill        & $2\times8\times5$& $2\times64\times5$
                                 & $2\times 96\times3$                       \\
$S_{\rm sys}$ (Jy)\dotfill     & 60               & 70          & 36         \\
$S_{\rm min}$ (mJy)\dotfill    & 1.2              & 1.0         & 0.15       \\
Pulsars found/detected\dotfill & 40/61            & 46/100      & 440+/630+  \\
Reference\dotfill              & Clifton et al. & Johnston et al. & this work\\
\tableline
\tableline
\end{tabular}
\end{table}

Despite this hindrance there are very good reasons for wanting to
search the Galactic plane: young pulsars will naturally be found close
to their places of birth, viz. the Galactic disk.  While relatively
rare, they are interesting for a variety of reasons, including the
study of pulsar--supernova remnant interactions, and the preferential
display of rotational ``glitches'', and increased likelihood of
emission at X- and $\gamma$-ray energies, which are of interest for
studies of the internal dynamics, and cooling and emission mechanisms
of magnetized neutron stars, respectively.  Also, to obtain an unbiased
picture of the intrinsic Galactic distribution of pulsars, rather than
just of the local population, one must penetrate deep into the Galaxy,
i.e., use high frequencies.

Only two large-area pulsar surveys had been carried out at 20\,cm prior
to the one described here.  The surveys of Clifton et al. (1992) and
Johnston et al. (1992) in the 1980s (see Table~1) were very successful
at finding many pulsars, preferentially young and relatively distant.

In early 1997 a 13-element receiver package with very good system noise
characteristics was installed on the Parkes telescope.  Developed for
surveying H{\sc i} in the local universe (Staveley-Smith et al. 1996),
it is also ideally suited for pulsar searching.

\section{Multibeam Survey}

We began collecting data for the survey in August 1997.  Receivers for
each of 13 beams are sensitive to two orthogonal linear polarizations.
Signals from each polarization of each beam are detected in 96 filters,
each 3\,MHz wide, upon which they are added in polarization pairs,
high-pass filtered with a cutoff of 0.2\,Hz, integrated for 0.25\,ms,
and 1-bit sampled before being written to magnetic tape with relevant
telescope information, for off-line processing.  The sensitivity of the
survey to long-period pulsars, about 0.15\,mJy, is a factor of seven
better than the previous Parkes 20\,cm survey (Johnston et al. 1992),
and we are even more sensitive to short-period pulsars, owing to the
faster sample interval and narrower filters used (see Table~1).
Figure~1{\em a\/} shows the calculated sensitivity of the survey.  Note
that the figure does not take into account scattering:  according to it
a pulsar like the Crab, with period 33\,ms, but located across the
Galaxy, with $\mbox{DM}=1000$\,cm$^{-3}$\,pc, will be detectable with a
minimum flux density of about 0.6\,mJy.  As Figure~4{\em b\/} shows,
such a pulsar will most likely not be detectable as a pulsed source
almost regardless of strength.

The 13 beam patterns (each subtending a $14'$ diameter) are not
adjacent on the sky; rather, one central beam (with the best
sensitivity) is surrounded by a ring of six beams separated by two
beams-widths, surrounded in turn by a second ring separated by a
further two beam-widths.  We collect data by interleaving pointings on
a hexagonal grid, resulting in complete sky coverage with adjacent
beams overlapping at half-power points.  Each pointing covers an area
about 0.6\,deg$^2$, resulting in sky coverage at a rate of
1\,deg$^2$/hr, and the total survey area requires about 2700 individual
pointings.

\begin{figure}
\plotfiddle{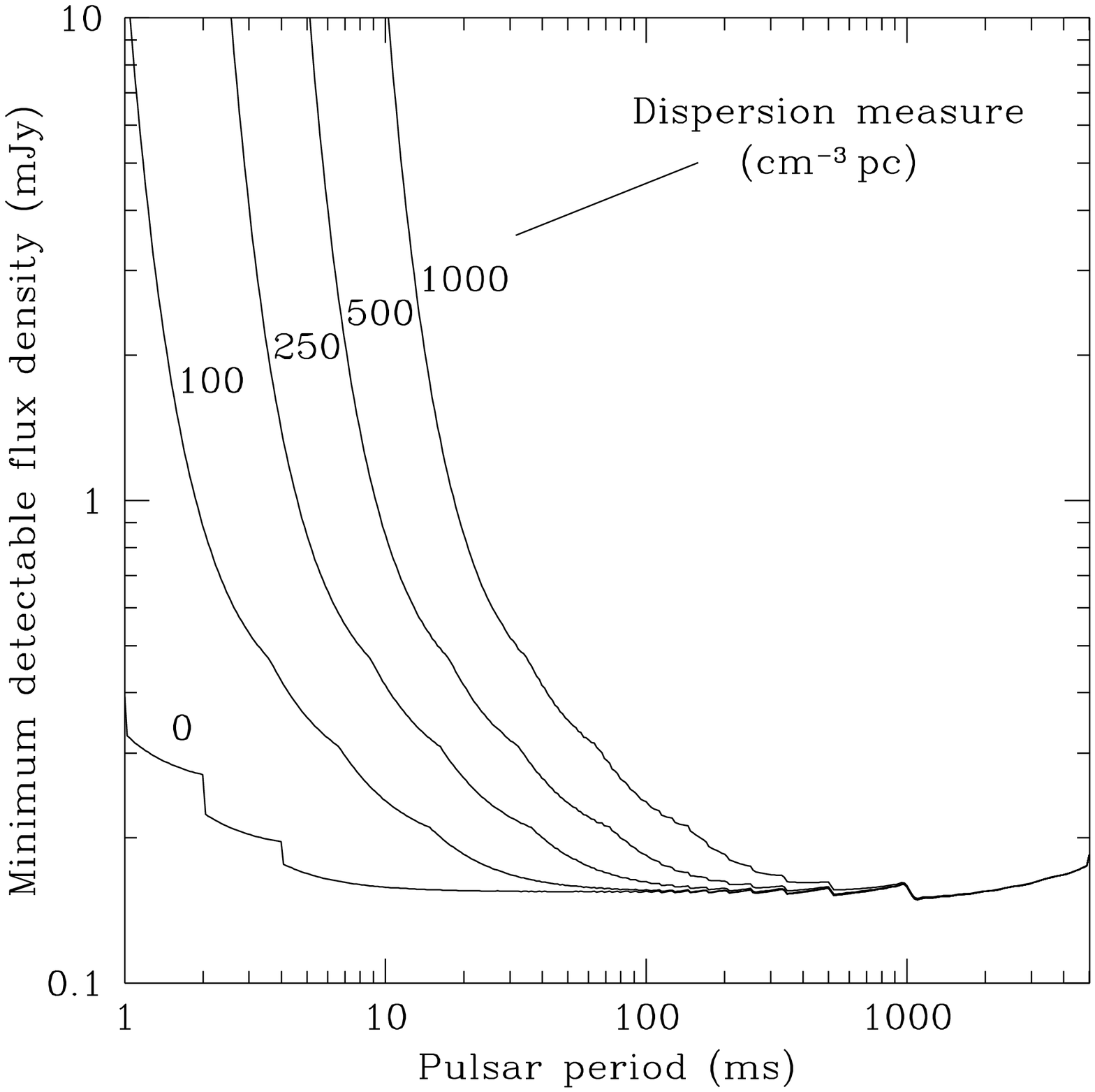}{5mm}{0}{29.5}{29.5}{-165}{-190}
\plotfiddle{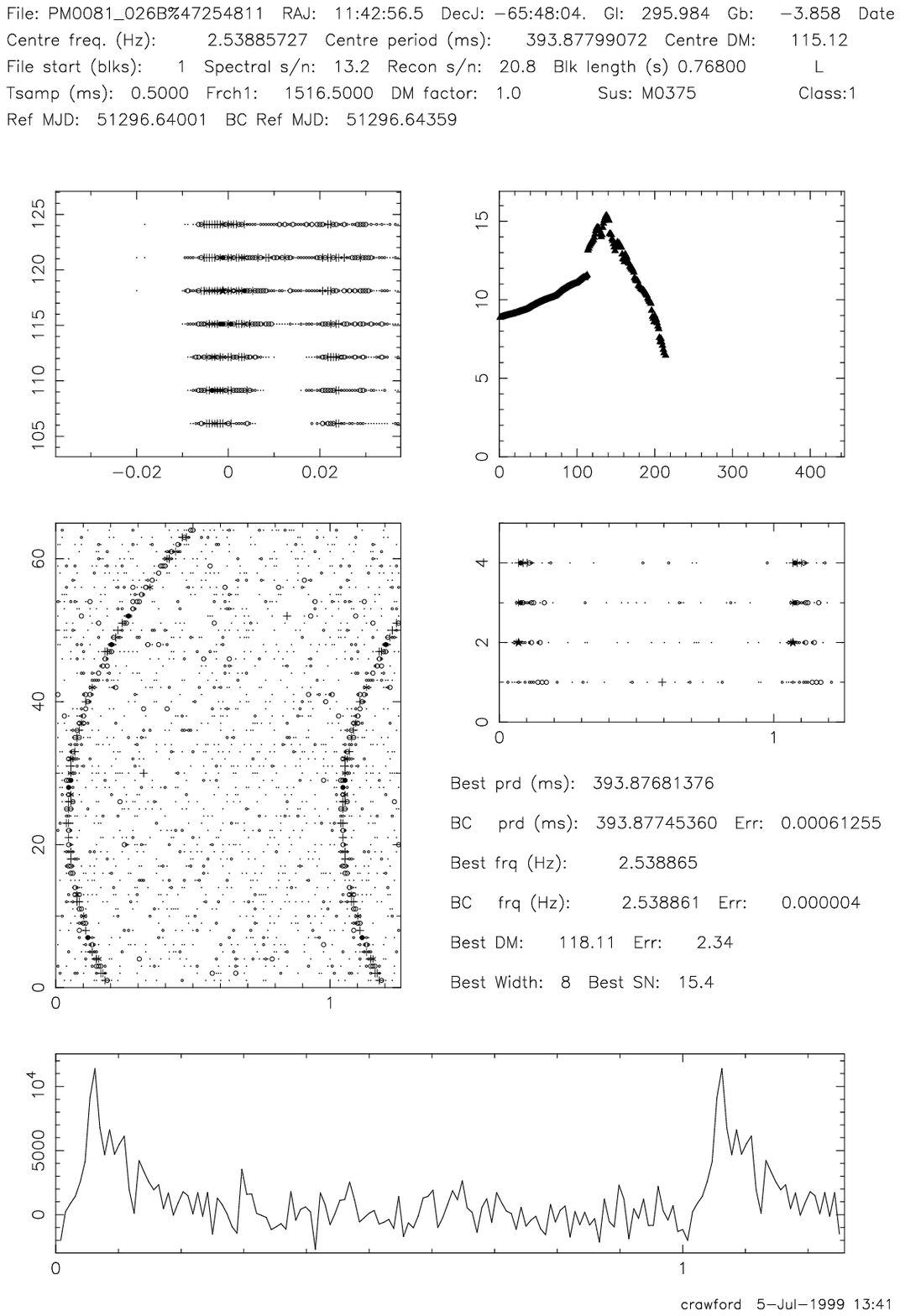}{5mm}{0}{36.5}{36.5}{-5}{-145}
\bigskip \bigskip \bigskip \bigskip \bigskip \bigskip
\bigskip \bigskip \bigskip \bigskip \smallskip
\caption{[{\em a\/}] Calculated survey sensitivity as a function of
period and DM for a pulsar observed at the center of a beam pattern.
[{\em b\/}] Output of search code for discovery of a binary pulsar in a
short-period orbit.}
\end{figure}

As of September 1999 we have collected and analyzed about 1200
independent telescope pointings, some 700\,deg$^2$, or 45\% of the
total.  Data reduction takes place in workstations in a manner similar
to previous surveys (e.g., Manchester et al. 1996).  Figure~1{\em b\/}
shows the search-code output for the discovery of a binary pulsar.  To
date we have discovered 440 new pulsars, and have re-detected 190 known
pulsars.  Because of the long integrations, some binary pulsars (in
particular millisecond pulsars) have signal-to-noise ratios reduced,
owing to Doppler-induced varying spin periods.  We therefore complement
our standard search analysis with ``acceleration search'' reduction,
recently begun.

\section{Discussion}

\begin{figure}
\plottwo{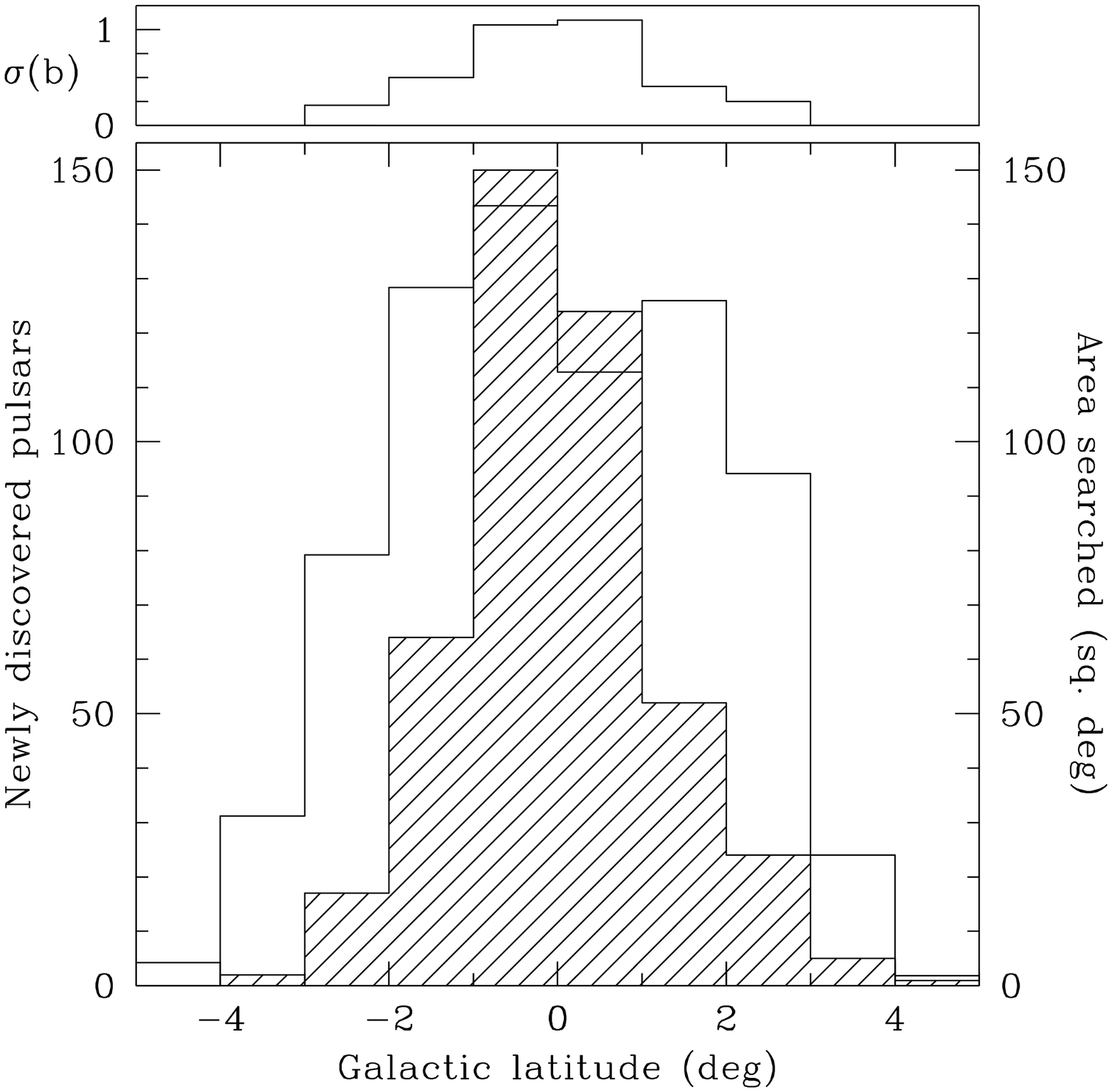}{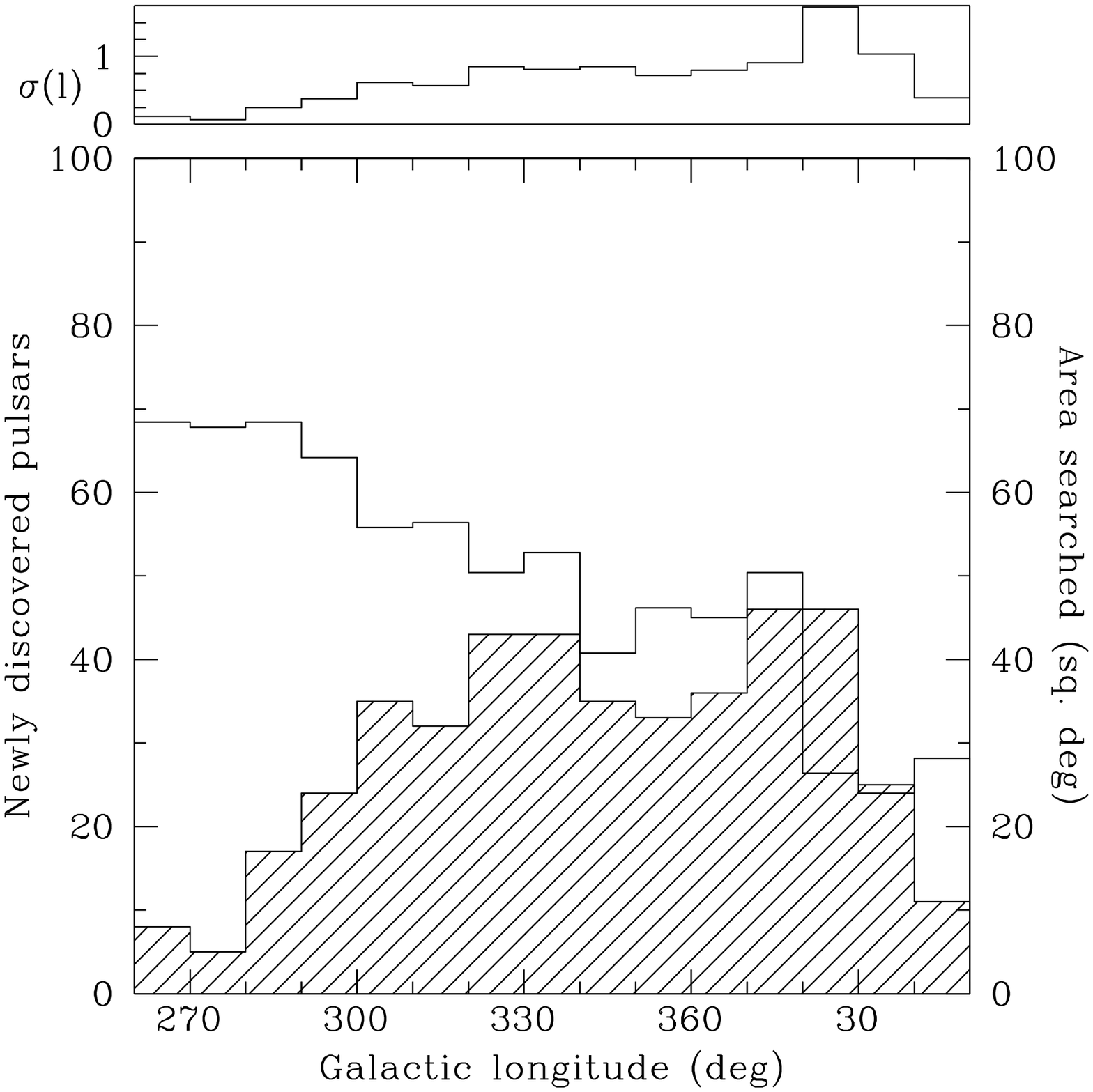}
\caption{[{\em a\/}] Pulsars discovered as a function of Galactic
latitude (cross-hatched histogram) and corresponding area searched so
far (line histogram).  The completed survey corresponds to having the
entire area of the figure searched.  At top is shown the density of
discoveries, in pulsars/square deg. [{\em b\/}] As for {\em a\/}, as a
function of Galactic longitude.}
\end{figure}

While we have discovered 10 new pulsars for every 1\% of the survey
area searched so far, it should not be concluded that we will amass a
total of 1000 new pulsars.  We began by surveying the regions closest
to the Galactic plane, which are richest in pulsars.  In Figure~2{\em
a\/} we see that we have already searched essentially the entire
$|b|<1^\circ$ region, with a new-pulsar density, averaged over the
entire longitude range, of slightly over 1/deg$^2$.  Clearly the
density of pulsars drops dramatically for $|b|\ga1^\circ$, which
comprises much of the region yet to be searched.  Conversely in
Figure~2{\em b\/} we see that, as a function of longitude, the region
we have preferentially searched so far ($260^\circ \la l \la
320^\circ$) has the lowest pulsar density.  Accounting for these
selection effects in some detail, we predict that the number of new
discoveries for the entire survey should be somewhat over 600.  Another
estimate is derived from the number of pulsars previously known in the
overall search area, 255, and the number of (re-)detections:  these
would suggest an eventual total of $(440/190)255 \simeq 600$ new
objects.  In fact we expect a total of about 700 pulsars to be
discovered, after accounting for some further selection effects (e.g.,
so far we have not been complete in confirming new pulsar candidates
down to the sensitivity level implicit in the calculations underlying
Figure~1{\em a\/}).  We have therefore already detected about 2/3 of
all pulsars we will discover.  Manchester et al. (this volume) describe
in some detail what is already known about many of the new pulsars.

Figure~3{\em a\/} shows the distribution of DM for the newly discovered
pulsars, and for comparison also shows that for previously known
pulsars.  Qualitatively the distribution is as expected: we find
pulsars predominantly with large DM, since our survey is by far the
most sensitive ever to distant pulsars along the Galactic plane; on the
other hand we find very few ``nearby'' ($\mbox{DM} \la
100$\,cm$^{-3}$\,pc) pulsars, in part because selection effects (e.g.,
scattering) preventing the detection of such pulsars in past surveys
were far less important than for high-DM pulsars.  The median DM for
the new pulsars is about 400\,cm$^{-3}$\,pc, and we also see that there
is a marked decrease in the number of objects with $\mbox{DM} \ga
900$\,cm$^{-3}$\,pc, with a maximum of about 1300\,cm$^{-3}$\,pc.
Naturally it is more difficult to find such highly dispersed pulsars,
both because of the attending level of scattering, and reduced flux
density due to large distance.  But as we can see from Figure~4{\em
a\/}, we do not in any case expect large numbers of pulsars in the
Galaxy to have $\mbox{DM} \ga 1400$\,cm$^{-3}$\,pc.

\begin{figure}
\plottwo{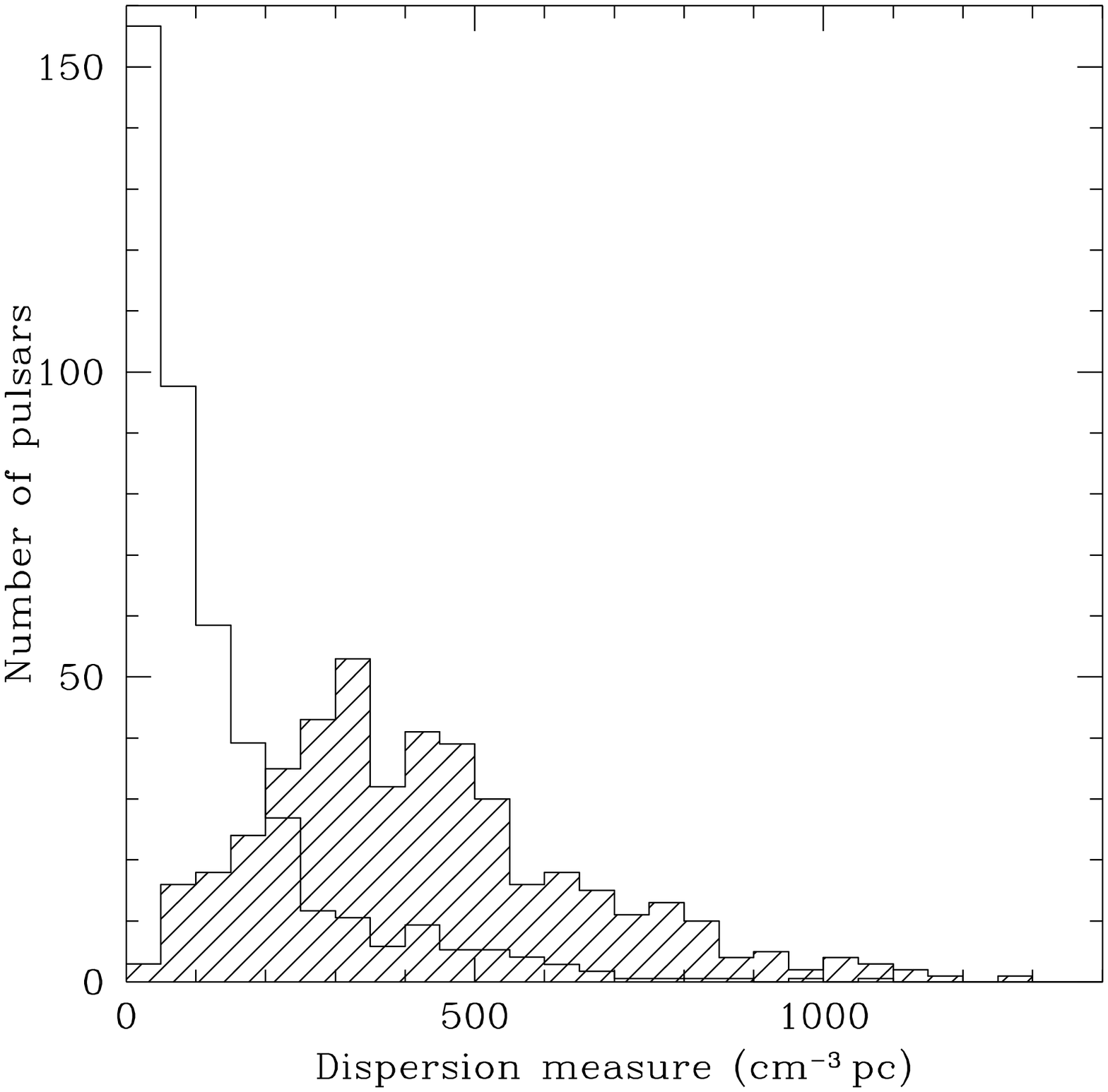}{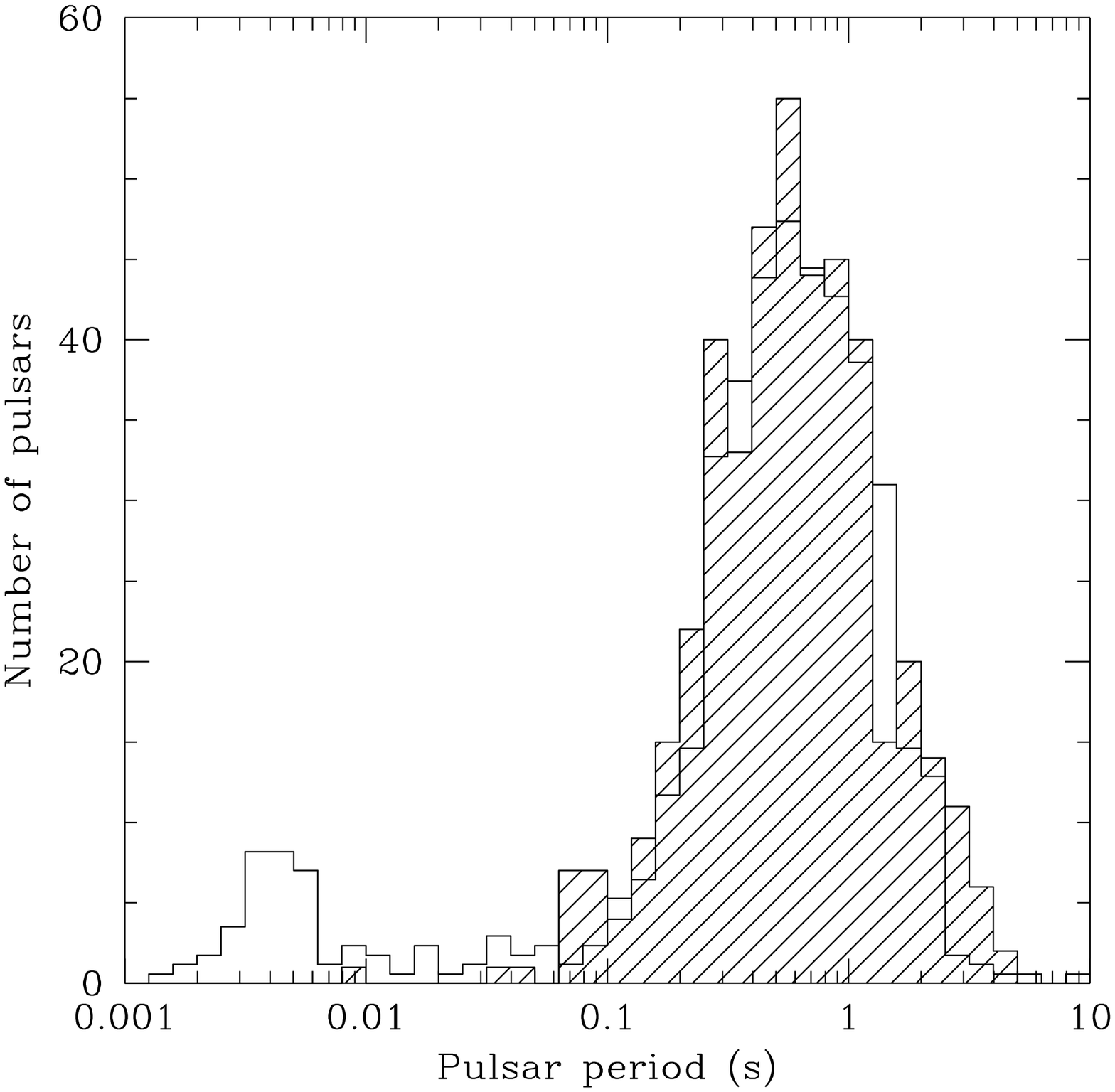}
\caption{[{\em a\/}] Histograms of DM for newly discovered
(cross-hatched) and previously known pulsars (line).  The area of the
latter histogram is scaled to equal that of the former.  [{\em b\/}]
Period histograms, as in {\em a\/}.}
\end{figure}

A comparison of the period distributions for newly discovered and
previously known pulsars (Fig.~3{\em b\/}) yields some surprises.  The
most obvious discrepancy is that we have not discovered a significant
number of short-period pulsars (only one with $P<30$\,ms).  The Parkes
70\,cm survey (Manchester et al. 1996), with a long-period sensitivity
of $\sim 3$\,mJy and poorer time- and frequency-resolution than the
multibeam survey, yielded 17 millisecond pulsars.  Assuming a spectral
index of $-2$, 3\,mJy at 70\,cm are equivalent to 0.3\,mJy at 20\,cm
--- i.e., we should be more sensitive to millisecond pulsars than that
survey.  One cannot reasonably appeal to hypothetical larger spectral
indices: Edwards et al. (this volume) report on the very successful use
of the multibeam data-acquisition system for finding millisecond
pulsars in a survey at intermediate Galactic latitudes and with short
integration times.  It is conceivable that strong scattering prevents
us from detecting many millisecond pulsars with $\mbox{DM} \ga
70$\,cm$^{-3}$\,pc; in any case we are still investigating this dearth
of millisecond pulsar discoveries.  A second surprise concerns the
discovery of relatively many pulsars with long periods, $P \ga 3$\,s.
This may be due to surveys with shorter integration times not
collecting data for many such pulse periods, coupled with the
prevalence of ``nulling'' pulsars among those with long periods; and to
the effectiveness with which much local radio-frequency interference
gets ``dispersed away'' in our search for high-DM pulsars.

In Figure~4{\em a\/} we plot the positions of the pulsars detected in
the multibeam survey projected onto the Galactic plane, according to
the distance model of Taylor \& Cordes (1993).  Two things are
immediately apparent: the previously known re-detected pulsars are
located relatively nearby ($d \la 4$\,kpc), while many of the newly
discovered pulsars are very distant, with many beyond the inner-most
spiral arm; and many of the newly discovered pulsars seem to be located
along spiral arms, particularly the two inner-most ones.  Regarding
this last point, we do expect pulsars to be correlated with spiral arm
locations, but we should be careful to consider potential biases in the
distance model that will place pulsars preferentially in regions of
high electron density.

\begin{figure}
\plotfiddle{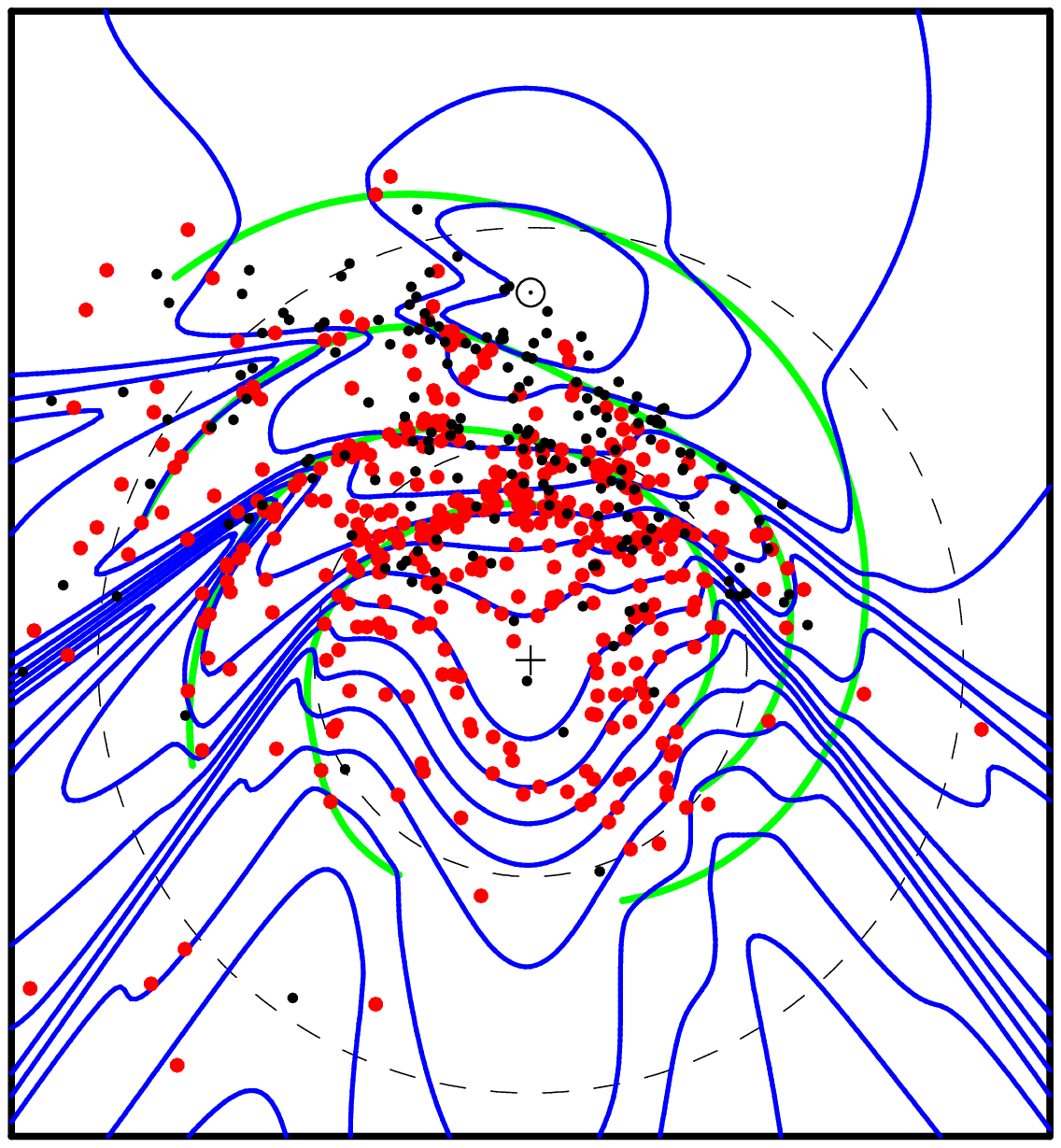}{5mm}{0}{51.5}{51.5}{-230}{-262}
\plotfiddle{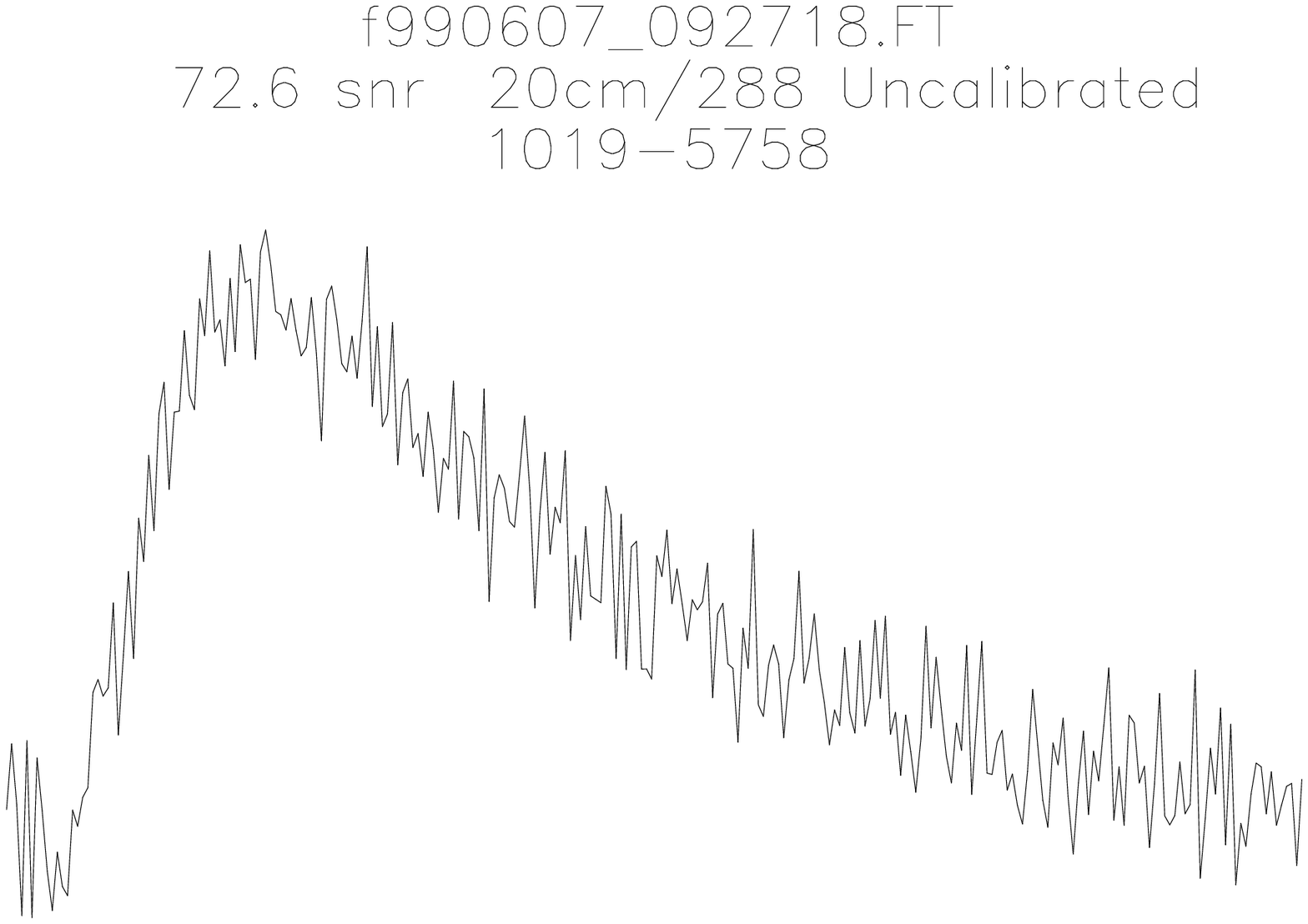}{5mm}{0}{34}{34}{-23}{-132}
\bigskip \bigskip \bigskip \bigskip \bigskip \bigskip
\bigskip \bigskip \bigskip \bigskip \smallskip
\caption{ [{\em a\/}] Heliocentric DM contours and pulsars detected.
The contours (first one for $\mbox{DM}=50$, next ones at
100\,cm$^{-3}$\,pc and multiples), and pulsar distances, are computed
from the Taylor \& Cordes (1993) free-electron model.  The Sun is
located at top ($\sun$), with the Galactic center (+) 8.5\,kpc away.
Four spiral arms are represented by grey lines.  Newly discovered and
previously known, but re-detected, pulsars are represented by grey and
smaller black dots, respectively.  [{\em b\/}]  Pulse profile at 20\,cm
of a newly discovered pulsar (with period 162\,ms and $\mbox{DM} =
1035$\,cm$^{-3}$\,pc) shows a significant ``scattering tail''.}
\end{figure}

The multibeam survey is being remarkably successful at uncovering very
distant pulsars.  The DM-distribution of these pulsars,
$\mbox{DM}(l,b)$, together with measured scattering parameters (cf.
Fig.~4{\em b\/}), and eventually Faraday-rotation parameters, probing
the interstellar magnetic field, should give us a more unbiased picture
of the Galactic distribution of pulsars, and add considerably to our
knowledge of the interstellar medium.

\end{document}